\documentstyle[paspkfo,twoside,psfig]{article}
\begin{document}

\title{ The determination of mass of stellar disks of galaxies from the kinematic data. }

\author{A.V. Zasov \altaffilmark{1}, A.V. Khoperskov \altaffilmark{2}, N.V. Tyurina \altaffilmark{1}}

\affil{Sternberg Astronomical Institute, 13, Universitetskii pr-t,
Moscow, Russia}

\affil{ Volgograd State University,20,2 Prodolnaya, Volgograd,
Russia}

\begin{abstract}

Different ways of the determination of masses of galactic disks,
based on the kinematic data, are briefly discussed. The analysis
of the rotation curves which reach maximum inside of a disk, and
N-body modeling, reproducing the rotation curves and stellar
velocity dispersion of real galaxies, enable to conclude that the
mass of a disk is usually significantly less than the total mass
of spherical components (bulge + dark halo) inside of optical
borders of a galaxy, although the exceptions also exist.

\end{abstract}

\section{Introduction}

 The determination of masses of components of galaxies
is hampered by the ignorance of the distribution of dark matter
(DM) within their optical borders: we do not know whether it is
concentrated in the disk (massive disk model) or in halo (massive
dark halo model) or in both.

The most direct way to decompose mass distribution into disk and
spherical components is the modeling of the rotation curves (RC)
accepting that the M/L ratio of stellar components is
approximately constant. In practice it usually means that the
scalelength of stellar density distribution is taken to be close
or equal to the photometric scalelength ${R_0}$. But even in this
approximation the results are usually ambiguous. The limiting
case which do not contradict to the observed rotation curves is
the "max disk model". In this case the contribution of stellar (+
gaseous) disk is scaled to explain the inner part of the RC,
assuming the the role of halo is small there. It should be
mentioned however that the result is rather sensitive to the
radial extension of the measured rotation curve and to its correct
shape. For the typical shapes of RC which have a plateau at large
R the resulting values of $M_{disk}/M_{halo}$ for "max disk model"
are close to 3 for R = 2$R_0$ and to 1 for R=4$R_0$. However the
reducing of mass of a disk down to factor 2-3 is usually also
compatible with the shape of RC.

Several independent methods were proposed to determine masses and
densities of disks, or to obtain some restrictions on the value of
ratio of masses of spherical to disk components $\mu$ of
individual galaxies, using the additional observational data.
Among them are:

    a) The presence or the absence of a bar ($\mu <1$ is the condition
    of the development of bar-mode instability in a dynamically cool
    disk (Ostriker, Peebles, 1973)). However it is evident that the
     relative mass of spherical component is not the only parameter
     which determine the condition of the existence of a bar,
     especially if to take into account that there are
    several physical mechanisms of bar formation and bar destruction.

    b) The introducing of spiral structure constraints
    related to the development of $m = 1$ and $m = 2$ modes, which imposes
    restrictions on the surface density of a disk (Athanassoula et
    al., 1987). A shortcoming of this approach is that it is based
    on a definite model of the origin of spiral density waves. In
    addition, the presence of a gas in the disk makes the
    condition of the large-scale density wave propagation much more complicated.

    c) The estimation of stellar velocity dispersion in the disk (by
    direct way or by the measurement of the disk thickness for edge-on galaxies).
    Indeed,
    gravitational instability "heats" a collisionless disk if its
    density exceeds some threshold value, which depends on the initial
    velocity dispersion and the RC. However there are no
    reliable analytical criteria of gravitational stability for 3-D
    collisionless disk, which makes the numerical modeling
    preferable (Bottema, 1993, 1997, Khoperskov et al., 2001,).

In this work we try to reveal the role of disk components using
some pecularities of the shapes of RCs of some galaxies and also
to determine the relative masses of spherical components $\mu$ for
concrete galaxies, using numerical experiments of dynamical
evolution of the disks.

\section {Galaxies with the local maximum on the rotation curve}
The shapes of the RCs of galaxies are rather different; it gives
evidence that their mass distribution is not identical. Some
fraction of rotation curves reach a maximum at some R=$R_m$
and/or decline in the periphery. The rotation curve component of
exponential disk passes through maximum at $R_m\approx 2.15 R_0$.
Therefore it is natural to expect that galaxies, in the inner
region of which the disk mass fraction is especially large, should
be among those where the observed RC has a maximum at $R_m \approx
2 R_0$. For these galaxies "max disk model" will give the most
justified solution, and, if the results are correct, the relative
M/L ratio for them should be systematically lower than for those
galaxies where a presence of DM smoothes out the maximum and
increases this ratio.
\begin{figure}[t]
\psfig{figure=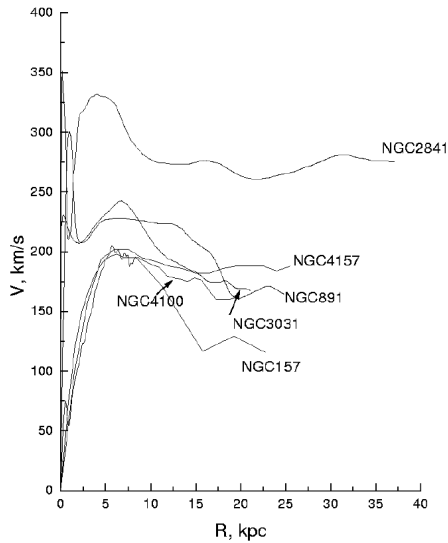,width=6 cm}
\caption{Rotation curves with
maxima at $R \approx 2R_0$}\label{Figure 1.}
\end{figure}

 To verify it, we extracted galaxies with the inner maxima of RC from two samples of rotation curves,
obtained with a good spatial resolution and presented by Sofue et
al., 1999 and Sanders $\&$ Verheijen, 1998. We also added to the
list of objects the galaxy
 NGC 157, which demonstrates well defined maximum of V(R) (Ryder et al.,
 1998) . Altogether we found 22 galaxies which are not members of
 strongly interacting systems, where V(R) has maxima  within
 the optical disk (circumnuclear extrema within R = 1 kpc were not considered).
 Practically all of these galaxies have photometrically determined values
 $R_0$. A comparison of $R_0$ and $R_m$ has shown that in most cases the
 maximum of V(R) has nothing to do with the stellar disks: only six
 out of 22 galaxies have  $R_m/R_0$ ratio close to 2. Their RCs are presented in Fig.1. We
 have modeled the RC of these galaxies using three components:  bulge, disk with the constant $M/L_B$ and
 the thickness 0.5 kpc, and spherical quasi-isothermal dark halo
 component.

\begin{table}
\caption {}
\label{Table1.}
\begin{center}\scriptsize
\begin{tabular}{lcr}
\tableline
Galaxy & Range of disk mass        & $M/L_B$      \\
       & ($10^{10}  M_{\odot}$)    & (solar units) \\
\tableline
NGC 157  &  4.9-5.2    & 1.6     \\
NGC 891  &  4.4 -11.6  & 11.1    \\
NGC 2841 &  7.6-11.5   & 10.0    \\
NGC 3031 &  3.9-5.2    & 6.0     \\
NGC 4100 &  3.1-5.6    & 6.5     \\
NGC 4157 &  4.6-5.6    & 7.6     \\
\tableline
\end{tabular}
\end{center}
\end{table}

A range  of disk mass values which satisfy the rotation curve of
chosen galaxies are given in Table1. The largest values
correspond to "max disk" model.
 With the exception of NGC 157, the range of possible
values of masses of disks remains rather wide. Curiously the total
mass within the optical diameter $D_{25}$ over luminosity $M/L_B$
is quite normal for these galaxies, being in the range 6-11 solar
 units (the only exception is NGC~157). It means that either the
 mass of DM within a given radius is quite normal for these galaxies
 in spite of the disk predominance in the inner $2R_0$, or the
 maximum of V(R) is caused not by the stellar disks, but by the
non-typical distribution of DM in the halo or by non-circular
motions.

The rotation curve of NGC 157 experiences a sharp drop just beyond
the optical radius (Ryder et al., 1998). It is the only galaxy we
found, where both signs of low mass of DM are present: first, the
very extended rotation curve  may be explained with low-massive
halo, and second, the integral $M/L_B$ ratio is unusually low.
\begin{figure}[t]

 \psfig{figure=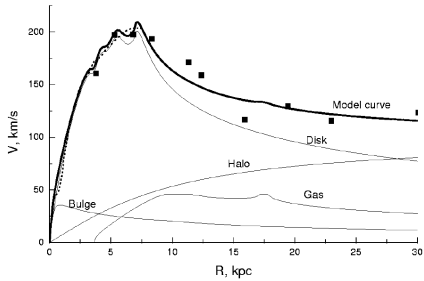,width=6cm }
\caption{Modeling of the rotation curve of NGC 157. Dashed line -
optical data, squares - HI data. } \label{Figure 2}
\end{figure}

\section {Numerical modeling}
A general idea we followed in  this work  is that the disks of
real galaxies should have the stellar velocity dispersion which
equals (or, in general case, exceeds) the minimal value necessary
for the disk to be gravitationally stable at a given R. To avoid
the problem of the absence of reliable analytical criterion of
stability for 3-D stellar disks, we use numerical simulations of
marginally unstable collisionless disk of a given galaxy, the
rotation curve and $R_0$ of which are known. At the end of
numerical experiments, the initially unstable disk heats up to a
steady condition and its kinematic properties ceased to change;
usually it takes place after about 5~-~10 periods of disk
revolution. Then we compared both the resulting model RC and
dispersion curve with the observed ones to choose the best fit
model.

\begin{figure}[t]

\psfig{figure=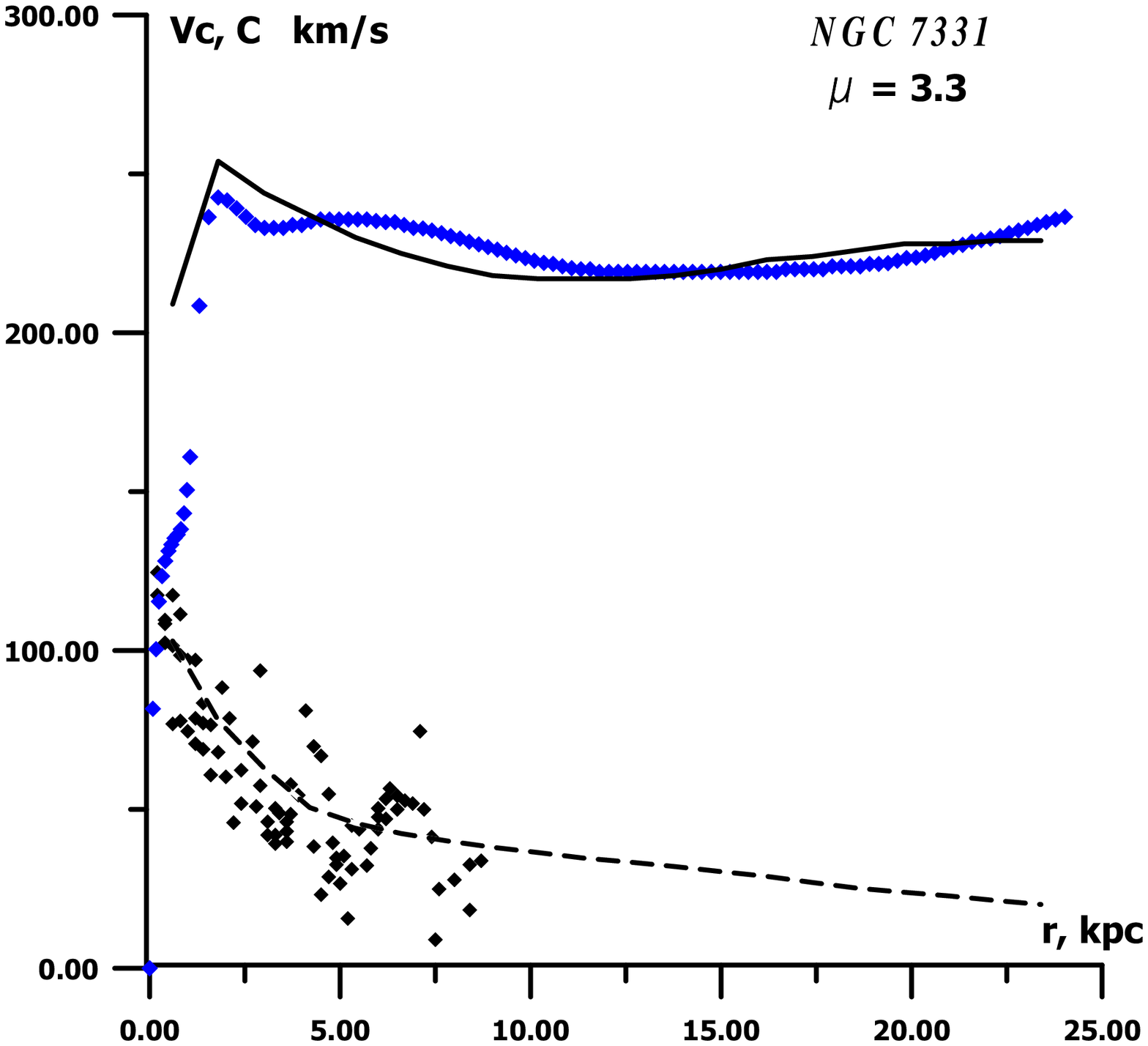,width=6 cm} \caption{A comparison of the
model and the observed velocities for NGC 7331. Thin line and
dashed line are circular RC and dispersion curve correspondingly
for the best fit model.}\label{Figure 3.}
\end{figure}

The details of N-body simulations are described by Khoperskov et
al., 2001. We used the direct p-p method to model the 3-D disk
with given values of scale length $R_0$ and the outer border at
$R=4R_0$, embedded into the rigid spherical components (bulge +
halo). The ratio $\mu$ of spherical to disk masses within
$R=4R_0$ was considered as a free parameter, which varied from one
model to another. Gas and young stars were not taken into account.

Initial velocity dispersion was chosen at sub-critical level for
gravitational instability. In practice it corresponds to Toomre
parameter $Q_T = ~C_r / C_T$ ($C_r$ is radial velocity dispersion
of stars,  $C_T ~=~ 3.36 \pi G \sigma / \kappa$ - Toomre' critical
velocity dispersion for axisymmetrical perturbations) $\approx$
0.8 - 1.2 along the radius. A set of experiments was done, which
showed that the end results are not critical to the choice of the
initial velocity dispersion if it remains subcritical and
bar-mode does not develop in a disk. In the disks which are cool
enough for the formation of a strong bar, the dynamical
evolution  of inner regions continues much longer and their
"heating" becomes more significant.

The experiments with different particle numbers from N = 2000 to
N = 80000 enable us to conclude that the end results practically
do not depend on  N, if $N > 15000-25000$ (depending on the
initial mass distribution in galaxies). However in the presence
of a bar the situation may be more complicated.

We applied N-body simulation to seven non-barred spiral galaxies
(see Table 2),
 for which the rotation curves, radial scale lengths $R_0$ and
radial distribution of stellar velocity dispersion of the disks
are known (Bottema, 1993, 1999, Corsini, 1999).

As one can expect, the resulting velocity dispersion at a fixed R
depends on $\mu$: it increases with the decreasing of relative
masses of spherical components. For self-gravitating disks
without spherical components $C_r$ is about 0.3-0.4 of maximal
circular velocity (Morozov, 1987, Khoperskov et al., 2001).
Line-of-sight components of the resulting model values of velocity
dispersion were compared with the observational measurements and
the best fit models were chosen. The example of model fitting for
NGC 7331 is illustrated in Figure 3.  The best fit values of $\mu$
and disk masses within $R=4R_0$ are given in Table 2.

\begin{table}
\caption{Parameters of modeled galaxies.} \label{Table2.}
\begin{center}\scriptsize
\begin{tabular}{lrrccc}

\tableline
Galaxy & Type &  D     & $r = 4R_0$ & $\mu $ &$M_{disk}$ \\
       &      &  Mpc   &    kpc     &        & $10^{10} M_{\odot}$\\
\tableline
NGC  891 &  SA(s)b      & 9.4   & 18.4 & 1.8 & 6.7 \\
NGC 1566 &  $(R'_1)SAB(rs)bc$       & 15.0  & 9    & 1.7 & 3.1 \\
NGC 2179 & (R)SA(r)0 & 35.6  & 16   & 1.1 & 4.9 \\
NGC 2775 & SA(r)ab   & 15.7  & 12   & 1.7 & 4.2 \\
NGC 3198 & SB(rs)c   & 8.8   & 10.4 & 2.1 & 1.8 \\
NGC 6503 &  SA(s)cd        &   5.9 & 4.6  & 2.0 & 0.5 \\
NGC 7331 & SA(s)b    & 14.9  & 24   & 3.3 & 6.9 \\
\tableline

\end{tabular}
\end{center}
\end{table}

\begin{figure}[t]
 \psfig{figure=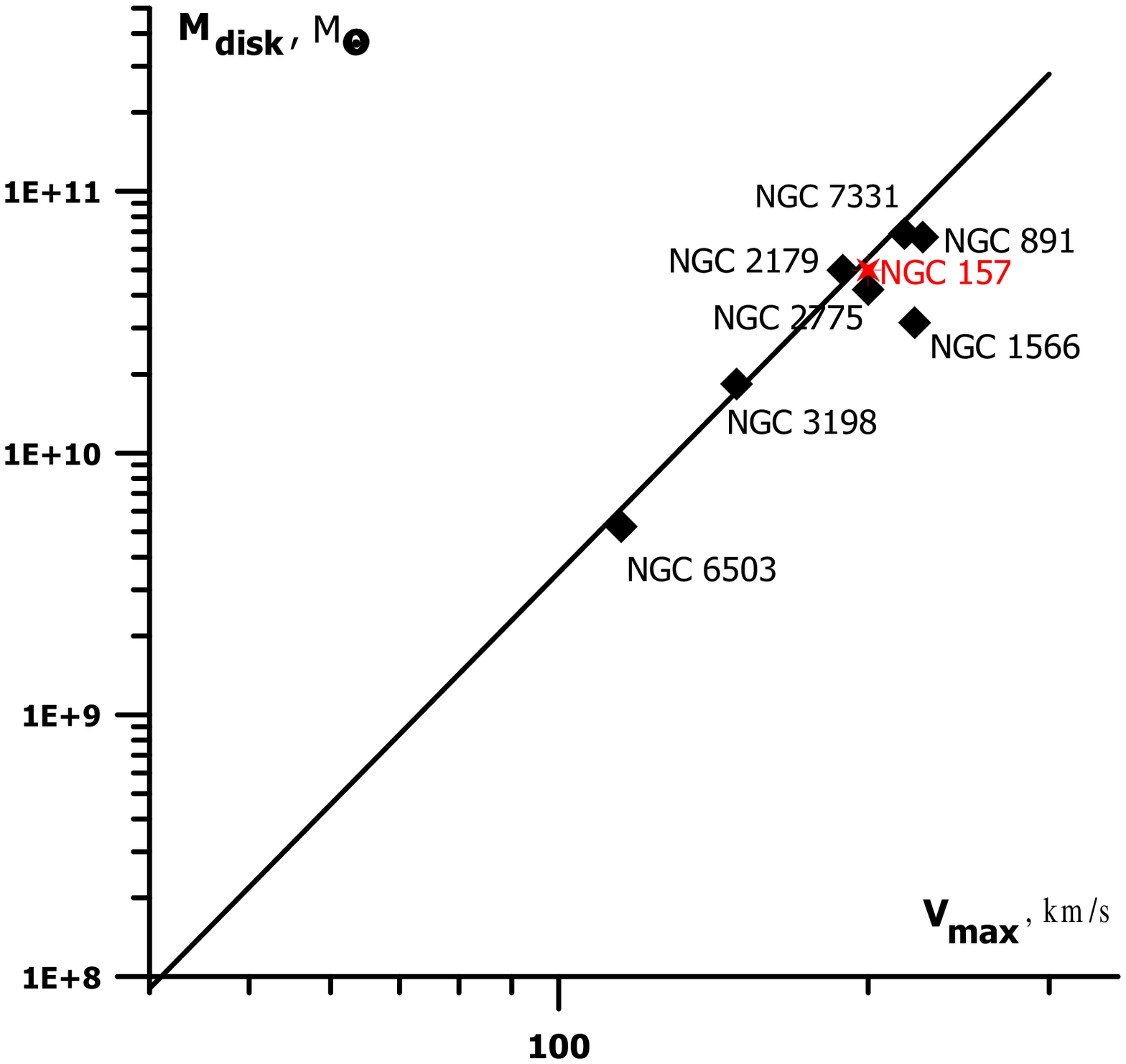,width=6 cm}
\caption{The dependence between $M_{disk}$ and maximal velocity
of rotation $V_{max}$.}\label{Figure 4.}
\end{figure}

\section {Conclusion}

In general, a comparison of numerical models with observations
favors the models with $\mu \geq 1.5$ within the radius $R=4R_0$
which exceed  the expected values of $\mu$ for the "max disk"
solution favoring the light disk model. Similar conclusions about
the predominance of heavy halo were discussed earlier by
 Bottema, 1993 (see also Khoperskov et al., 2001 and references therein).
 It is worth mentioning however that the dynamical heating of the disks may be
caused not only by the gravitational instabilities, but also by
some other reasons - for example by the encounter with neighbour
galaxies. In this case the model velocity dispersion should be
lower than the observed one for a given $\mu$. It means that the
model values of $\mu$ satisfying to the observations, are just a
lower limit, which makes the contrariety with the "max disk"
solution even more significant.

Figure 4 demonstrates that the mass of a disk $M_{disk}$ found for
galaxies discusses above is correlated with the maximal velocity
of a galaxy rotation $V_{max}$ (Tully-Fisher - like dependence).
This result both qualitatively and quantatively agrees with  the
$M_{disk} \sim V_{max}^4$ relationship (straight line), recently
found by McGaugh et al., 2000  for a sample of galaxies covering
more than five orders of magnitudes of masses. They used indirect
method of evaluation of $M_{disk}$, accepting $M/L_R$ = const for
stellar disks and adding mass of interstellar gas to it. The
origin of this relationship (and also Tully-Fisher relationship
between luminosity and $V_{max}$ of galaxies) is a problem which
has to be solved yet. It expresses the existence of a deep
connection between masses of visible disks and DM. The latter, as
it is confirmed in this work, is hidden in a halo rather than in
a disk and its mass is evidently preponderates over $M_{disk}$
within the optical borders of galaxies (however there some
exclusions also exist).

This work has been partically supported by the Russian Foundation for Basic Research,
grants 98-02-17102 and 99-07-90067.


\begin{references}


\reference Athanassoula E., Bosma A., Papaioannou 1987, \astap,
179, 23

\reference Bottema R. 1999, \astap...348...77B

\reference Bottema R. \& Gerritsen J.P.E. 1997, \mnras, 290, 585

\reference Corsini E. M., Pizzella A., Sarzi M. et al. 1999,
\astap, 342, 671

\reference Khoperskov A.V., Zasov A.V., Tyurina N.V. 2001, \arep,
in press

\reference  McGaugh S. S., Schombert J. M., Bothun G. D., de Blok
W. J. G, 2000, \apj, 533, L99

\reference Ostriker J.P. \& Peebles P.J.E., 1973, \apj, 184, 719

\reference Ryder S. D., Zasov A. V.,
 Sil'Chenko O. K.,
 McIntyre V. J., Walsh W. 1998, \mnras, 293, 411

\reference Sanders R. H. \& Verheijen M. A. W., 1998, \apj, 503,
97

\reference Sofue Y., Tutui Y., Honma M. et al., 1999, \astap, 523,
136


 \reference Zasov A.V., Mikhailova E.A., Makarov D.I.
1991, \alet, 17, 374


\end{references}
\end{document}